\documentclass[11pt]{article}
\usepackage{hyperref}
\usepackage{graphicx}
\renewcommand{\title}[1]{%
    \bigskip%
    \begin{center}%
    \Large\bf #1%
    \end{center}%
    \vskip .2in}

\renewcommand{\author}[1]{%
    {\begin{center}
    #1
    \end{center}}}
\newcommand{\address}[1]{\vspace{-1.7em}\vspace{0pt}
    {\begin{center}
    \it #1
    \end{center}}}
     
\begin{document}

\title{Elementary modes of coupled oscillators as whispering - gallery microresonators}

\author{Rabin Banerjee},
 
\address{S. N. Bose National Centre for Basic Sciences \\
JD Block, Sector III, Salt Lake City, Calcutta -700 098, India.}
 
\author{ Pradip Mukherjee}
 
\address{Barasat Government College\\
10 KNC Road, Barasat, Kolkata -700 124, India.
}

\begin{abstract}
{We obtain the elementary modes of a system of parity-time reversal ($PT$) - symmetric coupled oscillators with balanced loss and gain 
. These modes are used to give a physical picture of the phase transition recently reported \cite{BMG, NP1, NP2} in experiments with whispering - gallery microresonators.
 }\end{abstract}


\section{Introduction} 

It is well known that hermitian operators have real eigenvalues and are therefore used to construct physically meaningful objects. However, the converse is not true. An operator having real eigenvalues need not necessarily be hermitian. This fact gained prominence over the last decade when it was realised that consistent quantisation is possible with nonhermitian hamiltonians if the theory is $PT$ symmetric \cite{B,B2, B1}. 
 $PT$ symmetric theories are not merely of academic  interest. They have appeared in the study of experimental systems in diverse fields e.g
in superconductivity, quantum optics etc  \cite {NP2,NP3,NP4} and very recently, in optical microcavities. 

Optical microcavities have been the subject of many experimental and theoretical research in recent times. An important class of such systems has been branded as the whispering- gallery microresonators
(WGMR) \cite {N1}. Very recently an interesting experiment has been performed with a coupled system of WGMRs having balanced loss and gain \cite {NP1, NP2}. The importance of the experiment is related to  the fact that it implements a classical analogue of quantum systems described by non hermitian $PT$ symmetric hamiltonians. The experiment also demonstrates phase transition from 
$PT$ symmetry breaking phase to $PT$ symmetric phase. These results have been discussed theoretically by Bender et al \cite{BMG} who considered a coupled system of one-dimensional oscillators with balanced loss and gain. Specifically, it was shown that for a particular value of coupling, there is an onset of equilibrium between the oscillators.

Intense research over the last decade has shown that sensible quantum theories with non-hermitian hamiltonians can be constructed \cite{B, B1, BAM}. A new quantum theory of the damped harmonic oscillator (DHO) was constructed by the present authors
\cite{BM} which is based on the property $\eta^{-1}H^\dagger\eta = H$, where $ \eta = PT$ \footnote{ We can define the $\eta$ hermitian adjoint of a complex operator $A$ by $\tilde{A} = \eta^{-1}A^\dagger\eta$ and name it to be $\eta$ hermitian when $\tilde{A} = A$. The hamiltonians considered in \cite{BM} are $\eta$ hermitian.} The classical system discussed there consisted of the DHO and its time reversed image \cite{Ba, Ba1, Ba2}. Interestingly (though not accidentally), this doublet is just the system considered in \cite{BMG} in the limit of zero coupling. This suggests that similar techniques that were developed in \cite{BM} may be employed in the context of the coupled WGMR experiment \cite {NP1}. Taking another step forward, the elementary modes identified in \cite{BM} may be utilised to describe the interaction between the coupled oscillators. This  yields a new insight on the phase transitions mentioned in the beginning. This may facilitate the construction of a quantum theory of the coupled system.

In this paper we have provided an alternative approach to the problem providing novel results and fresh insights. We have explicitly constructed the lagrangian that corresponds to the coupled system of oscillators given in \cite{BMG}. This lagrangian is shown to be a composite of two $\eta$ hermitian model lagrangians. Such an analysis for zero coupling was earlier done by us \cite
{BM}. In that limit one of the elementary modes
 gives an exponentially growing solution and can thus be identified with the gain mode.
The other mode is the loss mode. As energy is pumped into the system the gain mode is excited and due to the coupling with the loss mode, energy dissipation takes place. The coupling results in mode mixing which changes the effective parameters of the elementary modes. The whole equilibrium process may thus be understood in terms of the energy sharing between the elementary modes. These modes correspond to the pair of microresonators in the WGMR experiment. The isolated modes are $\eta$ hermitian but not $PT$ - invariant. Only the composite system is $PT$ - symmetric. Thus until the coupling is stabilized the system is in 
$PT$ - symmetry breaking phase. The value of the coupling at which equilibrium is attained, or there is a phase transition from the broken $PT$ - symmetry to an unbroken one , is obtained from purely analytic means and reproduces earlier findings \cite{BMG}.

     The organaisation of the paper is as follows: in section II we present the whispering- gallery model \cite{BMG}. Some new results are given here pertaining to the issue of symmetry. In section III the elementary modes of the WGMRs are constructed. Using the soldering method geared for chiral oscillators \cite{BG1, BG2} it is shown that the composite model constructed from these basic modes is just the coupled model of \cite{BMG, NP1, NP2}. The equations of motion pertaining to the elementary modes have been written and their implications are discussed. Elements of the hamiltonian analysis are provided in this section.  In Section IV the  qualitative nature of the motion of the combined system is analysed which also explains the phase transitions. Section V contains our conclusions.

\section{The whispering- gallery model}
 We begin with a review of the model of \cite{BMG} which represents the coupled whispering- gallery microresonators (WGMR).
The equations of motion of the coupled system  are
\begin{eqnarray}
\ddot{x} - 2\gamma\dot{x} + \omega^2x  & = \epsilon y\nonumber\\
\ddot{y} +2 \gamma\dot{y} + \omega^2y &= \epsilon x\label{d1}
\end{eqnarray}
The physical system consists of two directly coupled microtoroidal whispering- gallery mode resonators \cite{N1} with balanced loss and gain \cite{NP1}. The first oscillator (the one with loss) is represented by $x$ whereas the second gain oscillator is represented by $y$. 
 The parameters $\gamma$ and $\omega$ are independent of time. Also observe that the same parameter $\gamma$ occurs
in both the equations signifying balanced loss and gain.

The special case $\epsilon = 0 $ is the uncoupled motion of the two oscillators that corresponds to Bateman's doublet consisting of a damped harmonic oscillator (DHO) and its time reversed image \cite{Ba, Ba1, Ba2} which has been analysed in a novel way in the recent past \cite{BM}. We will have more occasions to refer to this uncoupled limit. For now let us note that if
the ratio
\begin{equation}
R =\frac{\omega}{\gamma}\label{r}
\end{equation}
is greater than one,
the free motion of the DHO is oscillatory with exponentially decaying amplitude. Otherwise,
the motion is nonoscillatory i.e. overdamped \cite{BM}.

The Lagrangian of the system (\ref{d1}) can be constructed by the inverse lagrangian method \cite{S}.
First we write the variation of the action as
\begin{equation}
\delta S = \int^{t_2}_{t_1}dt\left[\left(\frac{d}{dt}\dot{x} 
         - 2\gamma\dot{x}      + \omega^2 x -\epsilon y\right)\delta y +
                \left(\frac{d}{dt}\dot{y} + 2\gamma\dot{y} + \omega^2 y -\epsilon x\right)\delta x\right]
                          \label{7}
\end{equation}
From (\ref{7}), the first equation of (\ref{d1}) is obtained by varying $S$ with respect to $y$
whereas the second equation of the set follows from varying $S$ with respect to $x$.
Since the equations of motion for $x$ and $y$ follow as Euler - Lagrange 
equations for $y$ and $x$ respectively, this method is called the indirect
method. Now, starting from (\ref{7}) we can deduce
\begin{equation}
\delta S = -\delta \int^{t_2}_{t_1}dt\left[\dot{x}\dot{y} - \gamma\left(x\dot{y} - \dot{x}y\right)
       - \omega^2xy +\frac{\epsilon}{2}(x^2 + y^2)\right]
                          \label{8}
\end{equation}
It is then possible to identify
the lagrangian of the system as
\begin{equation}
L = \dot{x}\dot{y} - \gamma\left(x\dot{y} - \dot{x}y\right)
       - \omega^2xy +\frac{\epsilon}{2}(x^2 + y^2) \label{n} 
\end{equation}
where $x$ coordinate and $y$ coordinate represent the loss and gain WGMRs. Either under parity $P$, 
\begin{equation}
t \to t\hskip .3cm x \to y \hskip .3cm {\rm {and}} \hskip .3cm y \to x \label{parity}
\end{equation}
or under time reversal
 $T$  which enforces 
\begin{equation}
t \to -t \hskip .3cm  x \to x \hskip .3cm {\rm {and}} \hskip .3cm y \to y \label{tr}
\end{equation}
the theory is not invariant.
However, invariance is achieved under the combined operation $PT$.

For the following analysis it will be advantageous to introduce the hyperbolic coordinates $x_1$ and $x_2$ \cite{BGPV} where,
\begin{equation}
x = {1\over \sqrt{2}}(x_1 + x_2); 
y = {1\over \sqrt{2}}(x_1 - x_2) \label{c3}
\end{equation}
 Using these  hyperbolic coordinates the Lagrangian (\ref{n})  can be written in a compact notation as
\begin{equation}
L = {1\over 2}g_{ij}\dot{x}_i\dot{ x}_j + \gamma\epsilon_{ij}x_i\dot{x_j} - { \omega^2\over 2}g_{ij}x_i x_j + { \epsilon\over 2}x_i x_i\label{L2}
\end{equation}
where the pseudo - Eucledian metric $g_{ij}$ is given by $g_{11}$ = -$g_{22}$ = 1 and $g_{12}$ = 0.
Under $PT$ 
\begin{equation}
x_i \to g_{ij}x_j \label{pt}
\end{equation}
Naturally, the  Lagrangian (\ref{L2}) is also $PT$ symmetric.

At this point one should note an important aspect of the problem. Without the coupling term (i.e when $\epsilon = 0$)
the bidimensional oscillator system (\ref{d1}) is invariant under the $SU(1,1)$ transformation
\begin{equation}
x_i \to x_i + \theta\sigma_{ij}x_j\label{88}
\end{equation}
where
$\sigma$ is the first Pauli matrix
and $\theta$ is an infinitesmal parameter.
The introduction of the coupling term breaks this invariance. The stronger  the coupling, more pronounced is the symmetry breaking. 

  Before finishing this section it will be appropriate to write the hamiltonian following from (\ref{L2}). It is given below
\begin{eqnarray}
H& =& {1\over{2}}\left(p_1 - {{\gamma}}x_2\right)^2\nonumber\\ 
 & +& \left({\omega^2 + \epsilon}\right)x_1^2 - \frac{1}{2}\left(p_2 + \gamma x_1\right)^2
                 -\left({\omega^2 - \epsilon}\right) x_2^2\label{I2}
\end{eqnarray}
where $p_i, i=1,2$ are the canonical momenta, conjugate to $x_i, i=1,2$, respectively. Note the difference of the effect of coupling in different degrees of freedom. When $\epsilon > \omega^2$ the last term changes sign signifying a qualitative change in the behaviour of the system.

\section{Elementary modes in the Lagrangian formalism}
The Lagrangian
(\ref{L2}) ( in conformity with (\ref{n}) ) is not invariant under parity. Thus the elementary modes of the WGMRs must be one dimensional yet planar oscllators with opposite sense of 'rotation'. These basic modes will now be explicitly constructed. They bear a striking resemblance to chiral oscillators that have  been utilised in the literature in connection with the Landau problem \cite{BG1,RL}, the DHO problem \cite{BM} etc. In fact the model (\ref{L2}) passes to the DHO problem in the limit $\epsilon \to 0$.
The resolution of the DHO in terms of complex lagrangians in the underdamped regime was of great use in the analysis of the model \cite{BM} . Similar reduction of (\ref{L2}) will now be discussed.
This will enable one to understand the mechanism of the interactions between the microresonators and will be useful for future investigations.

  Accordingly we introduce the Lagrangian doublet
\begin{equation}
L_{\pm} = \pm{\Gamma\over 2}\epsilon_{ij}x_i\dot{x}_j - {{k_{\pm}}\over 2}g_{ij}x_ix_j
\mp\frac{1}{2}\lambda x_ix_i\label{12}
\end{equation}
where, $\Gamma$, $k_{\pm}$ and $\lambda$ are as yet undetermined constants.
  The synthesis of $L_+$ and $L_-$ is now done by the soldering
formalism. Due to the presence of the constraints both $L_{\pm}$ have one degree of freedom each. They may be suitably combined to yield the lagrangian (\ref{L2}) which has two degrees of freedom.

We start from a simple sum
\begin{equation}
L(y,z) = L_+(y) + L_-(z)\label{s1}
\end{equation}
Substituting $L_{\pm}$ from (\ref{12})
we get 
\begin{equation}
L(y,z) = {\Gamma\over 2}\epsilon_{ij}y_i\dot{y}_j - {{k_{+}}\over 2}g_{ij}y_iy_j
-\frac{1}{2}\lambda y_iy_i - {\Gamma\over 2}\epsilon_{ij}z_i\dot{z}_j - {{k_{-}}\over 2}g_{ij}z_iz_j
+ \frac{1}{2}\lambda z_iz_i\label{s11}
\end{equation}
The essence of the soldering procedure can be understood  in the
following way. Use $x_i = y_i - z_i$ in $L(y,z)$ to eliminate
$z_i$ so that
\begin{eqnarray}
L(y,x) = &-&\frac{k_+}{2}g_{ij}y_i y_j -\frac{\Gamma}{2}\epsilon_{ij}
         \left[-2y_i\dot{x}_j  + x_i\dot{x}_j\right]\nonumber\\
        &-& \frac{k_-}{2}g_{ij}\left[y_i y_j - y_i x_j - x_i y_j + x_i x_j\right]
         + \frac{\lambda}{2}\left[-2y_i{x}_i  + x_i{x}_i\right]
        \label{sn5}
\end{eqnarray}
Since there is no kinetic term for $y_i$ it is really an auxiliary variable.
Eliminating $y_i$ from $L(y,x)$ by using its equation of motion
we directly arrive at 
\begin{equation}
 L(x) = -{{\Gamma^2}\over{2(k_+ + k_-)}}g
_{ij}\dot{x}_i\dot{x}_j - {{\Gamma(k_+ - k_-)}\over{2\left(k_+ + k_-\right)}}\epsilon_{ij}x_i\dot{x}_j - {{k_+k_-}\over{2(k_+ + k_-)}}g
_{ij}{x}_i{x}_j + \frac{\lambda\left(k_+ - k_-\right)}{2\left(k_+ + k_-\right)}x_ix_i\label{L2s}
\end{equation}
If we make the following identification
\begin{equation}
1= -{{\Gamma^2}\over{(k_+ + k_-)}},\hspace{.2cm}
\gamma= - {{\Gamma(k_+ - k_-)}\over{2\left(k_+ + k_-\right)}},\hspace{.2cm}
\omega^2 = {{k_+k_-}\over{(k_+ + k_-)}},\hspace{.5cm}\epsilon = \frac{\lambda\left(k_+ - k_-\right)}{\left(k_+ + k_-\right)}\label{id1}
\end{equation}
the lagrangian (\ref{L2s}) goes over to (\ref{L2}).
 Note that the opposite sign of the kinetic term of the elementary
lagrangians is crucial in the cancellation of the time derivative of $y$
in (\ref{sn5}) which in turn is instrumental in the success of the soldering
method.

The identification (\ref{id1}) has an immediate consequence. The ratio 
(\ref{r}) is found to be,
\begin{equation}
R =\frac{\omega}{\gamma}
= \left[1 - {{(k_+ + k_-)^2}\over{(k_+ - k_-)^2}}\right]^{\frac{1}{2}}\label{28}
\end{equation}
Different situations may arise depending on the coefficients $k_{\pm}$.
For real $k_+$, $k_-$, $R < 1$. Hence in this case the parameters identified by (\ref{id1})
correspond to an aperiodic motion of the free oscillators{\footnote{see the discussion
below (2)}}.
 Also note that for consistency of the first and third relations of (\ref{id1}) we
require $k_+$ and $k_-$ to be of opposite sign, with a suitable choice of their absolute 
values. Finally, $\Gamma > 0$ is required for positive $\gamma$,. 

We first consider the physically more important situation where the parameters of (\ref{L2}) must be such that the ratio $R > 1$. As already observed in the preceding paragraph this 
condition cannot be simulated by the identification (\ref{id1}) for real values of $k_{\pm}$. However, if $k_+$
and $k_-$ are continued to complex values so that
\begin {equation}
k_+ = \kappa\hspace{.5cm}k_- = \kappa^*\label{31}
\end{equation}
and
\begin{equation}
R = \left[1+\left({{Re\hspace{.1cm}\kappa}\over{Im\hspace{.1cm}\kappa}}\right)^2\right]^{\frac{1}{2}}\label{32}
\end{equation}
then clearly $R>1$, which is the required condition for oscillatory motion. Now equation (\ref{id1}) gives
\begin{equation}
1 = -{{\Gamma^2}\over{2Re\hspace{.1cm}\kappa}},\hspace{.5cm}\gamma = -{{i\Gamma Im\hspace{.1cm}\kappa}
\over{2Re\hspace{.1cm}\kappa}},\hspace{.5cm}\omega^2=
{{|\kappa|^2}\over{2Re\hspace{.1cm}\kappa}},\hspace{.5cm}\epsilon = +\frac{i\lambda Im\hspace{.1cm}\kappa}{Re\hspace{.1cm}\kappa}\label{33}
\end{equation}
We express $\kappa$ as 
\begin{equation}
\kappa = \kappa_1 +i\kappa_2\label{kn}
\end {equation}
 with $\kappa_{1,2}$ positive. The relations (\ref{33}) give
$\Gamma^2 = - 2\kappa_1$. We find that $\Gamma = \pm i \sqrt{2\kappa_1}$ i.e $\Gamma$ must be purely
imaginary. Again, using this value of $\Gamma$ in (\ref{33}) we get
\begin{equation}
\gamma = \pm \frac{\kappa_2}{\sqrt{2\kappa_1}},\hspace{.5cm}\lambda = -\left(\frac{i\kappa_1}{\kappa_2}\epsilon\right)\label{344}
\end{equation}
Note that there are two arbitrary real positive parameters $\kappa_1$ and $\kappa_2$ in the above. To simplify the expressions we put
\begin{equation}
\Gamma = \pm ig,\hspace{.4cm} g= \sqrt{2\kappa_1};\hspace{.4cm} \lambda = -i\Lambda\epsilon ,\hspace{.4cm}\Lambda=\frac{\kappa_1}{\kappa_2}\label{34}
\end{equation}
Here $\epsilon$ is kept explicit to show the effect of coupling.
Substituting (\ref{31}) and (\ref{34}) in  (\ref{12}) we get
 the elementary modes 
\begin{eqnarray}
L_+ = i\frac{g}{2}\epsilon_{ij}x_i \dot{x_j} - \frac{\kappa}{2}g_{ij}x_ix_j + i \frac{\Lambda\epsilon}{2}x_ix_i\label{ln1}\\
L_- = - i\frac{g}{2}\epsilon_{ij}x_i \dot{x_j} - \frac{\kappa^*}{2}g_{ij}x_ix_j - i\frac{\Lambda\epsilon}{2}x_ix_i\label{ln2}
\end{eqnarray}
the soldered form of which is the Lagrangian
 (\ref{L2}) pertaining to the oscillatory limit. Evidently $L_+$ and $L_-$ are a complex conjugated pair.

Several comments are due about the modes (\ref{ln1}, \ref{ln2}). First, they are first order planar lagrangians. There are two second class constraints which reduce the number of degrees of freedom by 2 in the phase space. Thus notwithstanding the fact that it is a two coordinate system, the number of degrees of freedom (in configuration space) is one. Next, these components satisfy  
\begin {equation}
L_{\pm} \to L_{\mp}\label{ref1}
\end {equation}
under $\eta = PT$. This can be easily checked by using equations (\ref{parity}, \ref{tr} and \ref{c3}). The result is consistent with the $PT$ - symmetry of (\ref{L2}).
In the uncoupled limit these elementary modes carry opposite Noether charges \cite{BM}. \footnote{ These can be compared with the dextro (right) and levo (left) rotatory modes of Fresnel construction in optics.} Since under $PT$-symmetry $x_1 \to x_1$ and  $x_2 \to -x_2$, the 'left handed' oscillator switches to the 'right handed' oscillator. The coupled motion may be viewed as interaction between
the elementary modes of opposite 'handedness'. 

 Now it will be useful to derive the equations of motion from the lagrangians $L_{\pm}$ given by (\ref{ln1}, \ref{ln2}). From $L_+$ we can easily find the equations corresponding to $x_1$ and $x_2$ respectively as
\begin{eqnarray}
 -ig\dot{x_2} +\left(k + i\Lambda\epsilon\right)x_1 = 0\label{x1}\nonumber\\
 ig\dot{x_1} - \left(k - i\Lambda\epsilon\right)x_2 = 0\label{x2}
\end{eqnarray}
From (\ref{x2}) we get
\begin {equation}
x_2 = \frac{ig\dot{x_1}}{k - i\Lambda\epsilon}\label{EOM11}
\end{equation}
Substituting this in (\ref{x1}) we get the equation of motion for $x_1$. Thus we find that
\begin {equation}
\ddot{x_i} +\frac{1}{g^2} \left( k^2 -\Lambda^2\epsilon^2 \right)x_i = 0, i = 1,2 \label{EOM1}
\end{equation}
Using the correspondence (\ref{33}) this can be written in terms of the physical parameters $\omega$ ,$\gamma$ and $\epsilon$ as
\begin {equation}
\ddot{x_i} + \left[\left(\Omega^2 -\gamma^2 + \frac{\epsilon^2}{4\gamma^2} \right) + 2i\Omega\gamma\right]x_i = 0, i = 1,2 \label{EOM2}
\end{equation}
Similarly, the equations of motion from the lagrangians $L_{-}$ are
\begin {equation}
\ddot{x_i} + \left[\left(\Omega^2 -\gamma^2 + \frac{\epsilon^2}{4\gamma^2} \right) - 2i\Omega\gamma\right]x_i = 0, i = 1,2 \label{EOM3}
\end{equation}
where
\begin {equation}
 \Omega^2 = \omega^2 - \gamma^2\label{X}
\end{equation}
Note that both (\ref{EOM2}) and (\ref{EOM3}) can be written in a compact way as
\begin {equation}
\ddot{x_i} + \Omega_{\pm}^2x_i=0, i = 1,2 \label{EOM33}
\end{equation}
where, 
\begin {equation}
\Omega_{\pm}^2 = \left[\left(\Omega^2 -\gamma^2 + \frac{\epsilon^2}{4\gamma^2} \right) \pm 2i\Omega\gamma\right]\label{W}
\end{equation}
For $\epsilon\ne 0$ it is not possible to write  the square root of the RHS of (\ref{W}) exactly as an elementary function of complex variables. Thus the physical interpretation of the oscillations in the general case is difficult to obtain. However, such an interpretation is manifest in the uncoupled limit.
Going over to the $\epsilon \to 0$ we find that $\Omega_{\pm} = \Omega\pm i\gamma$. One can then identify $L_+$ with the loss oscillator and $L_-$ with the gain oscillator. This observation will further be elaborated in the following section where an investigation of the coupled WGMRs is presented. 

 It will be appropriate to write the hamiltonians corresponding to the lagrangans $L_\pm$. From equation (\ref{ln1})) we can write $L_+$ as 
\begin{eqnarray}
L_+ = igx_1 \dot{x_2} - \frac{\kappa}{2}\left(x_1^2 - x_2^2\right)+ i \frac{\Lambda\epsilon}{2}
\left(x_1^2 + x_2^2\right)\label{ln11}
\end{eqnarray}
which is already in the first order form. The hamiltonian can thus be read off from (\ref{ln11})
as 
\begin{eqnarray}
H_+ =  \frac{\kappa}{2}\left(x_1^2 - x_2^2\right)- i \frac{\Lambda\epsilon}{2}
\left(x_1^2 + x_2^2\right)\label{hn11}
\end{eqnarray}
The canonical momentum conjugate to $x_2$ is $igx_1$
After a canonical transformation (CT)
\begin{equation}
x_+ = \frac{g}{\sqrt{k}}x_1 \hspace{.3cm};\hspace{.3cm} p_+ = -i\sqrt{k}x_2
\end{equation}
this hamiltonian can be written as 
\begin{equation}
 H_+ = \frac{1}{2}\left(1 + \frac{i\Lambda\epsilon}{k}\right) p_+^2 + \frac{k\left(k -i\Lambda\epsilon \right)}{2g^2}x_+^2\label{hplus}
\end{equation}
Similarly,the hamiltonian corresponding to the lagrangian $L_-$ (see equation (\ref{ln2})) can be written as 
\begin{equation}
 H_-= \frac{1}{2}\left(1 - \frac{i\Lambda\epsilon}{k^*}\right) p_-^2 + \frac{k^*\left(k^* +i\Lambda\epsilon \right)}{2g^2}x_-^2\label{hminus}
\end{equation}
where $x_-$ and $p_-$ are the complex conjugates of $x_+$ and $p_+$, respectively.

     It will be advantageous to express the hamiltonians (\ref{hplus},\ref{hminus}) in terms of the parameters $\gamma$, $\omega^2$ and $\epsilon$. Using the connections (\ref{34}) we get
\begin{eqnarray}
 H_+ &= &\frac{1}{2}\left[\left(1 + \frac{\epsilon}{2\omega^2}\right) + i \frac{\Omega\epsilon}{2\gamma\omega^2}\right] p_+^2 +\frac{1}{2}\left[\left(\Omega^2 - \gamma^2 + \frac{\epsilon}{2}
\right) + i \left(2\gamma\omega - \frac{\epsilon\Omega}{2\gamma}\right)\right]x_+^2\label{hplusn}\\
H_- &= &\frac{1}{2}\left[\left(1 + \frac{\epsilon}{2\omega^2}\right) - i\frac{\Omega\epsilon}{2\gamma\omega^2}\right] p_-^2 + \frac{1}{2}\left[\left(\Omega^2 - \gamma^2 + \frac{\epsilon}{2}
\right) - i \left(2\gamma\omega - \frac{\epsilon\Omega}{2\gamma}\right)\right]x_-^2\label{hminusn}
\end{eqnarray}
Further, invoking the CT
\begin{eqnarray}
 \pi_{\pm} &= &\left[\left(1 + \frac{\epsilon}{2\omega^2}\right) \pm i \frac{\Omega\epsilon}{2\gamma\omega^2}\right]^{\frac{1}{2}} p_{\pm}\label{c1}\\
X_{\pm} &= &\left[\left(1 + \frac{\epsilon}{2\omega^2}\right) \pm i \frac{\Omega\epsilon}{2\gamma\omega^2}\right]^{-\frac{1}{2}} x_{\pm}\label{c2}
\end{eqnarray}
we can express 
  \begin{eqnarray}
 H_{\pm} &= &\frac{1}{2} \pi_{\pm}^2 +\frac{1}{2}\left[\left(\Omega^2 - \gamma^2 + \frac{\epsilon^2}{4\gamma^2}
\right) \pm i 2\gamma\Omega 
\right]x_{\pm}^2\nonumber\\
 &= &\frac{1}{2} \left(\pi_{\pm}^2 +\Omega_{\pm}^2x_{\pm}^2\right)
\label{hminuspm}
\end{eqnarray}
where $\Omega_{\pm}^2$ is defined in (\ref{W}). This reproduces the expected form of the hamiltonians of the modes with the frequency $\Omega_{\pm}$, respectively .

\section{Driven motion of the coupled system -- the equilibrium conditions}
The motion of the coupled system (\ref{d1}) has been analysed in their elementary modes in the above section.
In the present section we will consider a situation where the coupled system is driven externally. This situation pertains to the coupled WGMRs experiment \cite{NP1}. Indeed, the elementary modes (\ref{ln1},\ref{ln2}) which are $\eta$ hermitian in isolation, will be identified as the coupled microcavities of the WGMR exeriment. Our aim will be to find the conditions for equilibrium from a physical point of view, in terms of the elementary modes.

\begin{figure}[ht]
\begin{center}
\includegraphics[width=6cm, height=5cm, angle=0]{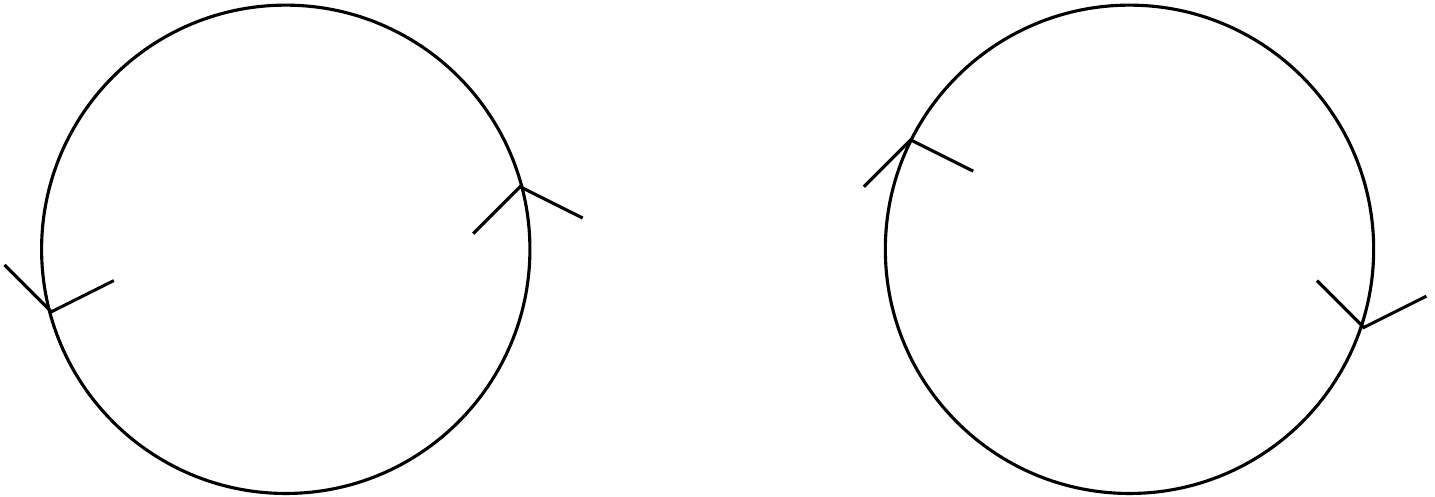}
\end{center}
\caption{{{The $PT$ breaking phase: the system behaves as two isolated chiral oscillators which are not $PT$ symmetric}} }\label{UC_plots}
\end{figure}

 To understand the physical situation it will be useful to study the uncoupled case ($\epsilon$ = 0). From equation (\ref{EOM2}) we see that in this case the complex frequency is given by            $\omega_+ $,
 where,
 $\ddot{x_i} + \omega_+^2 x_i = 0$
and $  \omega_+= \Omega + i\gamma$. 
The corresponsing solution for $x_1$ in the complex form is $x_1 = A\exp{i\left (\Omega + i\gamma\right)t}$. Again using $\epsilon$ = 0 in (\ref{EOM11}) and the relations $\Omega = \frac{k_1}{g}$ and 
$\gamma = \frac{k_2}{g}$ we get $x_2 = - A\exp{i\left (\Omega + i\gamma\right)}t $. Taking the real parts we get
 \begin{eqnarray}
 x_1 = A\exp{\left(-\gamma t\right)}\cos\left({\Omega t}\right)\nonumber\\ 
 x_2 = A\exp{\left(-\gamma t\right)}\sin\left({\Omega t -\frac{\pi}{2}}\right)
\label{sol1}
\end{eqnarray}
Together they form a right handed mode with loss. Similarly the mode corresponding to $L_-$ is a left handed mode with gain since its solution involves ($+\gamma$) instead of($-\gamma$) . Remember that these oscillatory modes are obtained when $\omega \ge \gamma$ (see equation (\ref{r}) and the discussion below it). The limiting condition for the underdamped motion can be written in terms of the complex frequency as 
\begin{equation}
 |{\omega_{\pm}}| = \gamma\label{AA1}
\end{equation}
If this condition is not realised the oscillatory gain mode is not excited. We assume this to happen for zero coupling. Thus the system of WGMRs behave as two oscillators in isolation. Since the individual oscilators are not $PT$ invariant, it is in the $PT$ breaking phase (see figure (\ref{UC_plots})).

\begin{figure}[ht]
\begin{center}
\includegraphics[width=6cm, height=5cm, angle=0]{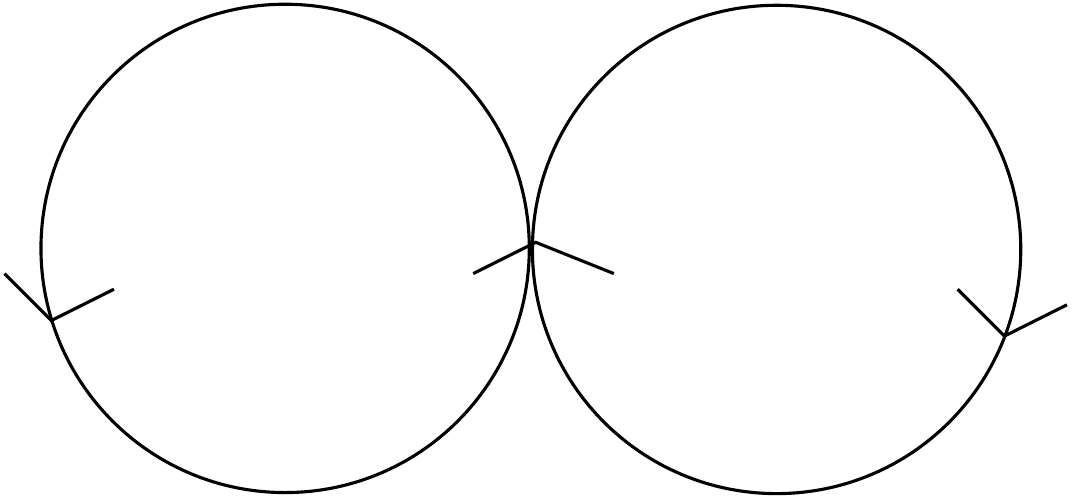}
\end{center}
\caption{{{Onset of $PT$ symmetric phase: the individual oscillators are not distinguishable.}} }\label{C_plots}
\end{figure}

The coupling is now initiated by bringing the WGMRs close enough. When the coupling is very small the motion of the gain oscillator is still non oscillatory and it cannot supply enough energy to the loss oscillator to establish stability. As the coupling is gradually increased a point comes when the motion of the gain oscillator begins. It then delivers the gained energy to the loss mode in a balanced way and equilibrium is attained (see figure (\ref{C_plots}). Unlike the uncoupled case we are unable to find the exact frequencies, but it is clear that the characteristic frequency will be modified by the coupling. Due to the feedback to the gain oscillator the damping will also be modified. However for small coupling we can neglect the effect on damping. Then, in analogy with (\ref{AA1}) we can define a condition that will dictate the onset of equilibrium. From (\ref{hminuspm}) and  (\ref{AA1}) this condition is given by $|\Omega_{\pm}|^2 = \gamma^2$ i.e.  
\begin{equation}
 \left(\Omega^2 -\gamma^2 + \frac{\epsilon^2}{4\gamma^2} \right)^2 + 4\Omega^2\gamma^2 = \gamma^4   \label{AA33}
\end{equation}
Squaring both sides and simplifying yields,
\begin{equation}
 \left[\omega^2 + \frac{\epsilon^2}{4\gamma^2} \right]^2 = \gamma^4 + \epsilon^2 \label{AA2}
\end{equation}
where, at an intermediate step, $\Omega$ has been eliminated in favour of $\omega$ by using 
(\ref{X}).
The solution for $\omega^2$ is now found by a perturbation expansion in powers of the coupling 
$\epsilon$,
\begin{equation}
 \omega^2 = \alpha_0 +  \alpha_1{\epsilon} + \alpha_2{\epsilon^2} + O\left(\epsilon^3\right) \label{p}
\end{equation}
subjected to the boundary condition that $\omega^2 = \gamma^2$ for $\epsilon = 0$. Substituting (\ref{p})in (\ref{AA2}) and retaining
terms upto $\epsilon^2$(which corresponds to the leading approximation), yields the condition,
 \begin{equation}
\epsilon^2 = 4\gamma^2\left(\omega^2 - \gamma^2\right)
\end{equation}
that implies the onset of equilibrium.
This condition was earlier obtained in \cite{BMG} by using a graphical technique to solve a quartic equation.

Now if the coupling strength $\epsilon$
is arbitrarily increased the point $\epsilon = \omega^2$ is reached. If $\epsilon$ exceeds this limit equation (\ref{I2}) shows that the coefficients of $x_1^2$ and $x_2^2$ have the same sign which qualitatively alters the structures of the oscillatory modes. The effective model is now bereft of the equilibration. The point $\epsilon = \omega^2$ thus gives an upper bound on the coupling strength for stable motion of the doublet.

\section{Conclusion} 
In this paper we have analysed a system of two coupled oscillators, one with  gain and the other with loss. For the particular case of identical gain and loss parameters the system becomes PT-symmetric and is closely connected with recent experiments \cite{NP1, NP2} on whispering- gallery microcavities, a fact that was first noticed in \cite{BMG}.

In the absence of coupling the present system consists of a damped harmonic oscillator and its time reversed image. A method was developed by us \cite{BM}, both in the lagrangian and hamiltonian formulations, to study such a model. Specifically the elementary modes of this composite model were shown to be chiral oscillators rotating in opposite directions. Chiral oscillators are basically simple harmonic oscillators with a preferred sense of rotation or `chirality'. The nontrivial ingredient now is the presence of coupling. However this fact was incorporated here and, with suitable modifications, we were able to construct the elementary modes. These modes were a pair of oscillators but their chirality gets smeared due to mode mixing brought about by the coupling.

A PT-symmetrical picture of the experiment on whispering- gallery microresonators emerged. In the absence of coupling, the elementary modes are the pair of chiral oscillators with opposite chirality, signifying loss and gain. These may be identified with the pair of microresonators with loss and gain. The individual lagrangians of the elementary modes are $\eta$ hermitian but not PT-symmetric. However, the composite system has $PT$ symmetry. Thus when the oscillators are isolated there is no PT-symmetry. As the coupling is switched on there is mode mixing between the basic oscillators. If the coupling is very small, the oscillatory condition of the gain mode is not attained so that there is no equilibrium. On increasing the coupling, there is a definite value when the oscillations commence and there is a balanced loss and gain. This signals the onset of equilibrium.

The particular coupling where equilibrium is attained was found by using analytic techniques and reproduced the results obtained by a graphical analysis \cite{BMG}. At this point the composite lagrangian (\ref{L2}) has PT-symmetry. This implies a phase transition from a broken PT- symmetric phase to an unbroken one. If the coupling is further increased, then there is a definite value where the oscillatory mode dies out and equilibrium is lost. This is the second phase transition point where an unbroken PT-symmetric phase passes over to a broken one. This second point  also agreed with previous results \cite{BMG}.

The success of our analysis depends on the Lagrangian method. If we have a coupled set of loss oscillators then it is not possible to use the inverse Lagrangian method to construct the Lagrangian of the system as we have done in (\ref{n}) from (\ref{8}). This is consistent with the problem of damped harmonic oscillator (DHO)  where we need to supplement the 
DHO by another oscillator which absorbs the dissipated energy \cite{Ba, Ba1, Ba2}. It may be interesting to see whether, by invoking the Bateman doublet for each of the loss oscillators, it is possible to form a conservative system and apply our method to find out the exceptional points.

To sum up, we were able to provide a physical picture of the WGMR experiments. It was based on a suitable extension of the method developed by us \cite{BM} to discuss uncoupled oscillators with balanced loss and gain. 
We also feel that the methods initiated in \cite{BM} and further developed here may be useful to analyse other PT-symmetry related phenomena (a good set of such references may be found in \cite{NP2}) based on chirality.


\newpage
\end{document}